\documentclass[12pt]{article}
\usepackage{a4wide,epsfig,amsmath,amssymb,cite,scalefnt}
\usepackage{color}

\newcommand{\abbrev}{\scalefont{.9}}
\newcommand{\drbar}{$\overline{\mbox{\abbrev DR}}$}
\newcommand{\msbar}{$\overline{\mbox{\abbrev MS}}$}

\newcommand{\msbarmath}{\overline{\rm\abbrev MS}}

\definecolor{NZGreen}{RGB}{0,225,0}

\definecolor{DKred}{RGB}{255,0,0}

\newcommand{\figurepath}{}

\begin{document}



\title{\vskip-3cm{\baselineskip14pt
    \begin{flushleft}
      \normalsize SFB/CPP-14-61 \\
     \normalsize TTP/TTP14-025 \\
     \normalsize Alberta Thy 14-14
  \end{flushleft}}
  \vskip1.5cm
${\cal O}(\alpha_s^2)$ corrections to the running top-Yukawa
  coupling and the mass of the lightest Higgs boson in the MSSM}
\author{ \small 
  D.~Kunz$^{(a)}$, L.~Mihaila$^{(a)}$, N.~Zerf$^{(b)}$\\
  {\small\it (a)  Institut f{\"u}r Theoretische Teilchenphysik,
 Karlsruhe Institute of Technology (KIT),}\\
  {\small\it 76128 Karlsruhe, Germany}\\
  {\small\it  (b) Department of Physics, 
   University of Alberta,}\\
  {\small\it 4-181 CCIS, Edmonton AB T6G 2E1, Canada}\\
 {\small\it E-mail: david.kunz@kit.edu, luminita-nicoletta.mihaila@kit.edu, zerf@ualberta.ca}\\
}

\date{}

\maketitle

\thispagestyle{empty}

\begin{abstract}
In this paper we propose a method to compute the running top-Yukawa coupling
in supersymmetric models with heavy mass spectrum  based on the ``running''
and ``decoupling'' procedure. In order to enable this approach 
we compute the two-loop SUSY-QCD radiative corrections required in the decoupling process. 
The method has the advantage that large logarithmic corrections are automatically resummed through the Renormalization Group Equations. 
As phenomenological application we study the effects of this approach on the
SUSY-QCD corrections to the prediction of the
lightest Higgs boson mass at three-loop accuracy. We observe a significant
reduction of the renormalization scale dependence as compared to the direct
method, that is based on the conversion relation between the running and pole mass for the
top quark. The effect of resummation of large logarithmic contributions
consists in an increased prediction for the Higgs boson mass, an observation
in agreement with the previous analyses. 


\end{abstract}


\section{Introduction}
The recent discovery of the Higgs boson at the Large Hadron Collider (LHC) was a  
major milestone   not just in particle physics but in the history of
science. It bears the knowledge how the mass comes about at quantum level by
the  Higgs mechanism. With the Higgs boson discovery, particle physics entered
a new era of tremendous 
intensity of detailed and careful study of  its properties. Hopefully, 
accurate understanding  of the
Higgs  phenomenology together with new information from experiments at the LHC
will provide us a tool for exploring new physics.\\ 
Great interest is currently devoted to the study of the Higgs boson couplings
to the electroweak gauge bosons $W$ and $Z$ and to the top- and bottom-quarks or
tau-leptons. Deviations in these couplings could possibly be observed 
once the currently large uncertainties will be improved as part of the LHC
program and at a future Higgs factory. It has been shown  that the Higgs couplings 
will be sensitive to new physics at the  multi-TeV scale  once percent
level precision on coupling measurements will be achieved (for a recent review
see \cite{Englert:2014uua}).

The aim of this paper is to propose a method for the  computation of the top-Yukawa coupling
within the Minimal Supersymmetric Standard Model (MSSM) taking into account
radiative corrections at ${\cal
  O}(\alpha_s^2)$ accuracy. In general, the  
 relationships between the running  couplings and the physical observables
like particle masses are affected by large radiative
corrections~\cite{Sirlin:1985ux,Hempfling:1994ar}. Within the SM, the 
relationship between  the running top-Yukawa coupling and the physical top-quark
pole mass  receives three type of radiative corrections:  i) higher order
corrections to the running Yukawa 
coupling, ii) contribution to
the fermion pole mass, and
iii) corrections to the relation between the Fermi constant and the SM
parameters. The first contribution is known at three-loop
accuracy~\cite{Chetyrkin:2012rz,Bednyakov:2012en} taking into account
corrections from all sectors of the SM. 
The radiative corrections to  the top-quark pole mass are known in QCD up to three-loop
order~\cite{Chetyrkin:1999ys,Chetyrkin:1999qi,Melnikov:2000qh,Marquard:2007uj}
and in the electroweak sector up to
two loops~\cite{Faisst:2003px,Jegerlehner:2003sp,Jegerlehner:2003py,Faisst:2004gn,Kniehl:2014yia,Eiras:2005yt}.
The third contribution  is known in the SM with
two- and three-loop
accuracy for the genuine electroweak~\cite{Fanchiotti:1992tu,Awramik:2002wn,Faisst:2003px} and
mixed QCD-electroweak~\cite{Chetyrkin:1995js} sectors, respectively.
 As was explicitly shown there are two sources for the  large radiative
 corrections: the
 pure QCD contributions to the top-quark pole mass of  about $10$~GeV, $2$~GeV
 and $0.5$~GeV~\cite{Chetyrkin:1999ys} at one-, two- and three-loop order, respectively; and  the
 tadpole diagrams  when they are taken into account for   a
 gauge-independent definition  of the running-mass~\cite{Faisst:2004gn}. Their
 magnitude is comparable with that from QCD sector  at
 the one-loop order and amounts to about $0.5$~GeV at  two loops.  As can be
 concluded from the numerical values cited above the radiative corrections are
 very important for the  QCD sector, and even the third-order terms
in the
 perturbative series are necessary in order to cope with the current
 experimental accuracy on the top-quark mass~\cite{ATLAS:2014wva}. 
However, the situation is
going to change if the International  Linear Collider (ILC) is built, where a
precision of ${\cal O} (100)$~MeV is expected.\\
When physics beyond the SM is considered, the radiative
corrections to the top-quark pole mass  might receive much larger contributions than in
the SM and even  diagrams beyond the three-loop order have to be taken into
account to reach the current experimental accuracy. By now, the radiative
corrections to the fermion pole mass are known at 
two-loop order for a general theory with massless gauge
bosons~\cite{Martin:2005ch}. The numerical evaluation of the two-loop
self-energy diagrams has been implemented in the code {\tt
  TSIL}~\cite{Martin:2005qm}. For supersymmetric models with masses 
at the few TeV scale~\footnote{In accordance with the lower bound from the direct
  SUSY searches at the LHC (for a review see~\cite{Craig:2013cxa}).} the radiative corrections to the
top-quark pole mass increase by about a factor of four as compared to the
SM results as we show in section~\ref{sec:numerics}. 
Thus, the two-loop contributions might become one order of magnitude
larger than the experimental uncertainties and the effects of higher order corrections have
to be considered. The computation of on-shell self-energy diagrams with
several mass scales  at the three-loop order  is, for the
time being, feasible only  using asymptotic expansion techniques and is
computationally very involved. In this paper, we propose an alternative method
 that can be applied as long as the top-quark mass is much smaller than the
 masses of supersymmetric particles. In the present paper, we focus on the
 dominant SUSY-QCD corrections to the running Yukawa coupling in the MSSM and
 postpone the study of the contributions originating from Yukawa and electroweak
 interactions for a later publication. The SUSY-QCD
 corrections to the relation between the running Yukawa coupling and the
 top-quark  pole mass reduce to the corrections to the ratio between the
 pole and the running masses. Explicitly, the running top-quark mass
 is determined  in the SM with the highest available  precision   from the
 experimentally measured pole mass~\footnote{For a detailed discussion about
   the distinction between the pole and the measured top-quark mass we refer
   to~\cite{Moch:2014tta}.}.  Then, the 
 running mass is evolved up to 
 the SUSY scale using the Renormalization Group Equations (RGEs) of  the
 SM. Afterwards, the running top-quark
 mass in the SM is converted to its value in the MSSM. In this step, threshold corrections 
at the SUSY-scale are required. In the last step, the running top-quark mass
in the MSSM is evolved at the desired energy scale with the help of MSSM RGEs.
As the RGEs in the SM~\cite{Chetyrkin:2012rz,Bednyakov:2012en} and the
MSSM~\cite{Ferreira:1996ug,Harlander:2009mn} are known to three-loop order,
the threshold corrections are 
required at the two-loop order. They are known for light quark  masses (e.g. bottom
quark) in the SM to three-loop
accuracy~\cite{Chetyrkin:1997un,Steinhauser:1998cm} and in the MSSM to
two loops~\cite{Bednyakov:2007vm,Bauer:2008bj}. Great interest was devoted to
the determination of the effective bottom-Yukawa coupling in SUSY-models with a large
$\tan \beta$ parameter~\cite{Carena:1999py,Mihaila:2010mp,Noth:2010jy}.  For
these models  the resummation of $\tan \beta$ enhanced contributions is considered
 on top of the  two-loop order calculation.  

One goal of this paper is
to present the computation of the two-loop SUSY-QCD 
threshold corrections to the running top-quark mass. The calculation is similar
with the ones performed in~\cite{Bednyakov:2007vm,Bauer:2008bj}. The advantage
of the method presented here as compared to the direct calculation of on-shell
self-energy diagrams in the  MSSM is that the occurring large logarithms of the form
$\ln(M_{\rm top}/M_{\rm SUSY})$ are automatically resummed by the use of
RGEs. The  result is a much better convergence of the perturbative series as
will be explicitly shown in the next sections.\\
The second
aim is to study the phenomenological effects of the above
calculation. Obviously, the top-Yukawa coupling is an essential ingredient in
all processes involving interactions between the Higgs boson and top-quark and
top-squarks. However, the most prominent example is probably the effects on  the
lightest Higgs boson mass, that receives radiative corrections enhanced by the
fourth power in top-Yukawa coupling. A detailed numerical analysis of the
effects of the proposed method for the determination of the running Yukawa
coupling on the SUSY-QCD corrections to the lightest Higgs boson mass will be discussed in section~\ref{sect:mhphen}.

The paper is structured as follows: in section~\ref{sec:frm} we present our
computational framework; explicit analytical results are discussed in
section~\ref{sec:res} together with their numerical implementation; in section~\ref{sect:mhphen}
we perform a phenomenological analysis of the effects of the two-loop SUSY-QCD
corrections determined in the previous section on the SUSY-QCD prediction of the
lightest Higgs boson mass with three-loop accuracy; in section~\ref{sec:concl}
we summarise our conclusions.


\section{\label{sec:frm} Framework}
An elegant approach to get rid of  unwanted large  logarithms occurring in the
predictions for observables in  multi-scale processes 
 is to formulate an effective theory (ET) (for a  review  see Ref.~\cite{Steinhauser:2002rq}).
 The parameters of the ET must be modified  in
order to take into account the effects of the heavy fields. The ET parameters are related to the
parameters of the full theory by the so-called matching or decoupling
relations. These take into account threshold corrections generated by the heavy
degrees of freedom that are integrated out  when the ET is constructed. In
the following, we concentrate on the calculation of the decoupling
coefficients for the strong 
coupling and the top-quark mass within SUSY-QCD. They are defined through the
following relations between the bare quantities\\
\begin{eqnarray}
\alpha_s^{0,\prime}&=&\zeta_{\alpha_s}^0 \alpha_s^0\nonumber\\
m_t^{0,\prime}&=& \zeta_{m_t}^0 m_{t}^0\,,
\end{eqnarray}
where the  primes label the quantities in the effective theory.
The decoupling coefficients have been computed in QCD including
 corrections up to the four-loop
order  for the strong coupling~\cite{Schroder:2005hy,Chetyrkin:2005ia} and 
three-loop order for quark masses~\cite{Chetyrkin:1997un,Steinhauser:1998cm}.
In the MSSM the  two-loop 
 SUSY-QCD~\cite{Harlander:2005wm, Bednyakov:2007vm,Bauer:2008bj}
 and SUSY-EW~\cite{Bednyakov:2007vm,Noth:2010jy} expressions are  known. Very
 recently, even the three-loop SUSY-QCD corrections to decoupling
 coefficient of the strong coupling were computed~\cite{Kurz:2012ff}.

We consider SUSY-QCD with $n_f=6$ active quark and squark flavours and $n_{\tilde{g}}= 1$ 
gluinos. Furthermore, we assume that  all SUSY-particles including squarks and the gluino are much
heavier than the SM particles. Integrating out the heavy
fields from the full  Lagrange density, we obtain the Lagrange density corresponding to
the ``effective'' SM  with $n_f$  quarks plus non-renormalizable interactions. The latter are
suppressed by negative powers of the heavy masses and will be neglected here.\\
Since the decoupling coefficients are universal quantities, they are
independent of the momenta carried by the incoming and outgoing
particles. The authors of 
Refs.~\cite{Chetyrkin:1997un}  showed
 that the  bare decoupling coefficients for the quark mass
$\zeta_m^0$  and for the strong coupling constant $\zeta_s^0$ can be  derived 
via the relations 
\begin{eqnarray}
\zeta_{\alpha_s}^0 &=&
\left[\frac{1+\Gamma^{0,h}_{gcc}(0,0)}{(1+\Pi^{0,h}_c(0))\sqrt{1+\Pi^{0,h}_g(0)}}\right]^{\!2}\,,\nonumber\\
\zeta_{m_t}^0 &=& \frac{1-\Sigma_s(0)}{\sqrt{\zeta_L
    \zeta_R}}\quad\mbox{with}\nonumber\\
 \zeta_L&=& 1+\Sigma_v(0)-\Sigma_A(0)\quad\mbox{and }\quad \zeta_R\,\,=\,\,1+\Sigma_v(0)+\Sigma_A(0)\,,
\end{eqnarray}
where $\Sigma_s(p^2)$, $\Sigma_v(p^2)$ and $\Sigma_A(p^2)$ are the scalar, the
vector and the axial-vector components of the top-quark self-energy defined
through
\begin{eqnarray}
\Sigma(p^2)= /\!\!\! p (\Sigma_v(p^2) +\gamma_5 \Sigma_A(p^2)) +m_t \Sigma_s(p^2)\,.
\end{eqnarray}

\begin{figure}
  \begin{center}
      \epsfig{figure=\figurepath 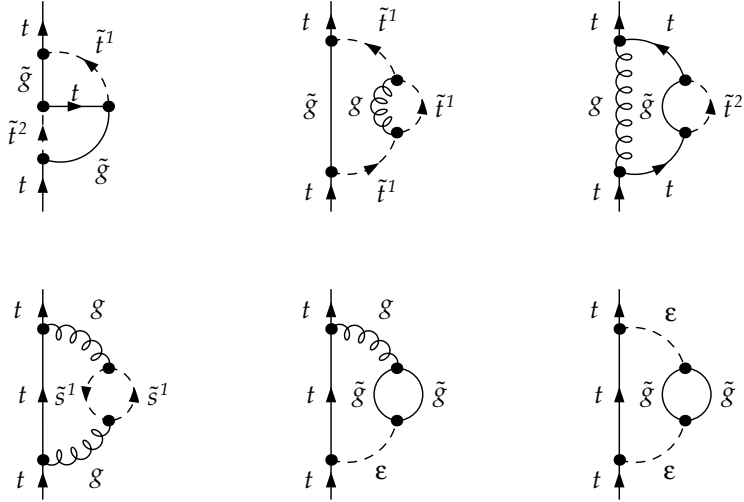,width=.61\textwidth}
  \end{center}
  
  \caption[]{\label{fig::diagrams} Sample Feynman diagrams that
    contribute to the two-loop corrections to $\zeta_{m_t}$.          
    }
\end{figure}

Note that the axial-vector component starts contributing at the two-loop
order. $\Pi^{0,h}_c(p^2)$ and $\Pi^{0,h}_g(p^2)$ are the ghost and gluon vacuum
polarizations and $\Gamma^{0,h}_{gcc}(p,q)$ denotes the amputated Green function
contributing to the gluon-ghost-ghost vertex.
  The
 superscript $h$ indicates that in the framework of Dimensional Regularization
 (DREG) or Dimensional Reduction (DRED) only diagrams
 containing at least one heavy particle inside the loops contribute and that
  only the hard regions in the asymptotic expansion of the diagrams are
  taken into account.\\
In  Fig.~\ref{fig::diagrams}
 some sample Feynman diagrams contributing to the
decoupling coefficients for the top-quark mass are shown.

The Feynman diagrams were generated with
{\tt QGRAF}~\cite{Nogueira:1991ex} 
and further processed with   {\tt q2e} and {\tt exp}
~\cite{Harlander:1997zb,Seidensticker:1999bb}. The 
reduction  of various 
vacuum integrals to the master
integral was performed with a self written {\tt
  FORM}~\cite{Vermaseren:2000nd} routine~\cite{js:2011}. The reduction of  
topologies with two different massive  and one massless lines
requires a careful treatment.  The related master integral
can be easily derived from its general expression  valid for massive
lines, given in
Ref.~\cite{Davydychev:1992mt}.

The finite decoupling coefficients are obtained upon the
renormalization of the bare parameters. They are
given by
\begin{eqnarray}
  \zeta_s = \frac{Z_s}{Z_s^\prime} \zeta_s^0
  \,,\quad  \zeta_m = \frac{Z_m}{Z_m^\prime} \zeta_m^0\,,
  \label{eq::zetagren}
\end{eqnarray}
where $Z_s^\prime$ and $Z_m^\prime$ correspond to the renormalization
constants in the 
effective theory, and  $Z_s$ and $Z_m$ denote the same quantities in the
full theory. Since we are interested in the two-loop results for
$\zeta_i\,,\, i=s,m$, the corresponding renormalization constants for
SUSY-QCD and QCD have to be implemented with the same accuracy. Analytical 
results for the latter up to the three-loop order can be found in  {\it e.g.} 
Refs.~\cite{ Jack:1994kd, Steinhauser:2002rq, Bednyakov:2002sf}.  

Apart from the renormalization constants of the external fields, also
parameter renormalization is required.  For
the renormalization of the gluino and 
squark masses  and the squark mixing angle we choose the on-shell
scheme.   This scheme allows us to directly use 
the physical parameters in the running analyses making the implementation
very simple. The explicit formulae at the one-loop order can be found in
Refs.~\cite{Pierce:1996zz,Harlander:2004tp}. The two-loop counterterms
are known analytically only for specific mass hierarchies~\cite{Kant:2010tf}
and numerically for arbitrary masses~\cite{Martin:2005qm}.\\
For the calculation of  $\zeta_{\alpha_s}$ the simultaneous  renormalization
of up- and down-type squarks is required. We follow the prescription proposed
in Ref.~\cite{Heinemeyer:2004xw} and fixed the counterterms for the up-squarks and the heavier
down-squark to be on-shell
and derive the counterterm for the lighter down-squark accordingly. In our
limit of neglecting the top-quark mass as compared to the SUSY mass scale it
holds
\begin{equation}
 \delta m _{\tilde{d}_1}= \delta m _{\tilde{u}_1}^{\rm
   os}\,,\quad\mbox{with}\quad u=u,s,t\quad\mbox{and}\quad d=d,c,b\,,
\end{equation}
where $\delta m _{\tilde{d}_1}$ stands for the mass counterterm of the
light down-squark and $\delta m _{\tilde{u}_1}^{\rm os}$ denotes the on-shell
mass
counterterm of the light up-squark.

For the computation of the decoupling coefficient of the top-quark
mass at order ${\cal O}(\alpha_s^2)$ one needs to renormalize in
addition the top-quark mass and
 the $\epsilon$-scalar mass. 
As  the top-quark mass is neglected w.r.t. SUSY particle masses, an
explicit dependence of the radiative corrections on $m_t$ can occur only
through top-Yukawa coupling. In order to avoid the occurrence of
large logarithms of the form 
$\alpha_s^2\log(\mu^2/m_t^2)$ with $\mu\simeq \tilde{M}$, where $\tilde{M}$
stands for the SUSY scale, one has to
renormalize  
the top-Yukawa coupling in the \drbar{} scheme. In this way, the large
logarithms are absorbed into the running mass and the higher
order corrections are maintained small. As a consequence, the top-squark
mixing parameter $X_t=A_t-\mu_{\rm SUSY}/\tan\beta$ will be renormalized in a
mixed scheme. Its counterterm is derived from the  relation
between the top-quark and squark masses and the mixing angle and  mixing parameter.

Finally, the last parameter to be renormalized is the $\epsilon$-scalar
mass~\footnote{For a review on their role in multi-loop calculation
  see~\cite{Mihaila:2013wma}.}. To obtain decoupling coefficients 
independent of the unphysical parameter $m_\epsilon$, one has to
modify the top squark
masses by finite quantities~\cite{Jack:1994rk,Martin:2001vx}. We adopt in the
present calculation the method proposed in Ref.~\cite{Bednyakov:2007vm} to
choose the $\epsilon$-scalars massive and integrate them out together with the
SUSY particles. In this way we achieve  conversion from \msbar{}
to \drbar{} schemes and decoupling of the heavy masses in a single step.

\section{\label{sec:res} Two-loop threshold corrections}
The two-loop results for the decoupling coefficient for the strong coupling is
very compact and we reproduce them below.
\begin{align}
\zeta_{\alpha_s}&=-\frac{\alpha_s}{4\pi}
\frac{1}{3}\bigg[C_A\left(1+2 L_{\tilde{g}}\right)
  +T \sum_{\tilde{q}}\sum_{i=1,2} L_{\tilde{q}_i}\bigg]
 + \left(\frac{\alpha_s}{4\pi} \right)^2\bigg\{
C_A^2\left(-\frac{125}{18}-\frac{44}{9} L_{\tilde{g}}+\frac{4}{9}
L_{\tilde{g}}^2 \right)
\nonumber\displaybreak[1]\\
&
+ C_A T
\frac{2}{9}\bigg[30+\sum_{\tilde{q}}\sum_{i=1,2} \bigg(6\frac{
    M^2_{\tilde{q}_i}}{ M^2_{\tilde{g}}}
+6  \frac{ M^2_{\tilde{g}}-M^2_{\tilde{q}_i}}{ M^2_{\tilde{g}}} B_{0,{\rm fin}}(M^2_{\tilde{g}},M_{\tilde{q}_i},0)+2 L_{\tilde{q}_i} L_{\tilde{g}}
\nonumber\displaybreak[1]\\
&
 -\frac{2 M^4_{\tilde{g}}-5 M^2_{\tilde{g}}
  M^2_{\tilde{q}_i}+6 M^4_{\tilde{q}_i}}{M^2_{\tilde{g}}(M^2_{\tilde{g}}-M^2_{\tilde{q}_i})}L_{\tilde{q}_i}
+3 \frac{M^2_{\tilde{g}}}{M^2_{\tilde{g}}-M^2_{\tilde{q}_i}}L_{\tilde{g}}\bigg)
\bigg]
+ T^2\left(\frac{1}{3} \sum_{\tilde{q}}\sum_{i=1,2} L_{\tilde{q}_i}\right)^2
\nonumber\displaybreak[1]\\
&
+ C_F T
\frac{2}{3}\bigg[+\sum_{\tilde{q}}\sum_{i=1,2}\bigg(1+\frac{M^2_{\tilde{g}}}{M^2_{\tilde{q}_i}}-\frac{M^2_{\tilde{g}}-M^2_{\tilde{q}_i}}{M^2_{\tilde{q}_i}}B_{0,{\rm
      fin}}(M^2_{\tilde{q}_i},M_{\tilde{g}},0)-2\frac{3 M^2_{\tilde{g}}-2 M^2_{\tilde{q}_i}}{M^2_{\tilde{g}}-M^2_{\tilde{q}_i}} L_{\tilde{q}_i}
\nonumber\displaybreak[1]\\
&
+\left(4+ \frac{M^2_{\tilde{g}}}{M^2_{\tilde{q}_i}}+\frac{2 M^2_{\tilde{q}_i}}{M^2_{\tilde{g}}- M^2_{\tilde{q}_i}}\right)L_{\tilde{g}}
\bigg)
+  \sum_{\rm gen}\bigg(-3\frac{M^2_{\tilde{q}_{u1}}}{M^2_{\tilde{q}_{d1}}}
-\frac{M^2_{\tilde{g}}-M^2_{\tilde{q}_{u1}}}{M^2_{\tilde{q}_{d1}}} B_{0,{\rm
    fin}}(M^2_{\tilde{q}_{u1}},M_{\tilde{g}},0)
\nonumber\\
&
+
\frac{M^2_{\tilde{g}}-M^2_{\tilde{q}_{d1}}}{M^2_{\tilde{q}_{d1}}} B_{0,{\rm
    fin}}(M^2_{\tilde{q}_{d1}},M_{\tilde{g}},0)+
L_{\tilde{q}_{d1}}-\frac{M^2_{\tilde{q}_{u1}}}{M^2_{\tilde{q}_{d1}}}
L_{\tilde{q}_{u1}} +\frac{1}{2} \frac{M^2_{\tilde{q}_{d1}}}{M^2_{\tilde{g}}-
  M^2_{\tilde{q}_i}} L_{\tilde{g}}
\bigg)
\bigg]
\bigg\}
\end{align}
In the formula above the sum $\sum_q$ runs over all quark flavours and
$\sum_{\rm gen}$ over the number of generations. $C_A,C_F$ are the quadratic Casimir invariants for the
adjoint and fundamental representations, $T=1/2$ is  the Dynkin
index and $L_x=\ln(\mu^2/m_x^2)$, with $x=\tilde{g},\tilde{q}$.
$B_{0,{\rm    fin}}(p^2, m_1,m_2)$ denotes the fine part of the $B_0$-function~\cite{Denner:1991kt}. The asymmetry
w.r.t. up- and down-type quarks originate from the special renormalization
scheme of down-type squarks relative to the up-type squarks. Here, $\alpha_s=\alpha_s^{\rm{ (SQCD)}}$
denotes the strong coupling constant in the full  theory.

\subsection{Limits}
The final results for two-loop threshold corrections for $\zeta_{m_t}$ are too lengthy to be
displayed here.  They are available in
{\tt Mathematica} and {\tt Fortran}
format from http://www.ttp.kit.edu/Progdata/ttp14/ttp14-025. Instead, we present them
 for two special mass hierarchies.
\subsubsection{Scenario A}
We consider first the case of all supersymmetric particles having masses
of the same order of magnitude and being much heavier than the top-quark.
\[
\begin{split}
&m_{\tilde{u}} = \ldots = m_{\tilde{b}} = m_{\tilde{t}} =
  m_{\tilde{g}}=\tilde{M}\quad \gg\quad 
m_t\nonumber\\ 
& \alpha_s^{(6)} = \zeta_{s}^{\tilde{M}}\, \alpha_s^{\rm{ (SQCD)}} \,,\qquad 
m_t^{(6)} = \zeta_{m_t}^{\tilde{M}}\, m_t^{\rm{ (SQCD)}}\,.
\end{split}
\]
$\zeta_{s}^{\tilde{M}},\,\zeta_{m_t}^{\tilde{M}} $ are functions of the
supersymmetric mass $\tilde{M}$, the
soft SUSY breaking
parameters $X_t=A_t-\mu_{\rm SUSY}/\tan\beta$,
the strong coupling constant  in the full theory $ \alpha_s^{\rm{ (SQCD)}}$,  and the
decoupling scale $\mu$. The superscript $(6)$ indicates that the
parameters are defined in QCD with $n_f=6$  quarks.
\begin{eqnarray}
\zeta_{m_t}^{\tilde{M}}&=&-\frac{\alpha_s}{4\pi} C_F \left(-1
+\frac{X_t}{\tilde{M}}+L_{\tilde{M}} \right) 
\nonumber\\
&&
+\left(\frac{\alpha_s}{4\pi}
\right)^2\bigg\{
C_F^2\bigg[-\frac{71}{8}-\frac{13}{2} L_{\tilde{M}}+\frac{1}{2}
  L_{\tilde{M}}^2 
+\frac{X_t}{\tilde{M}}\left(-5+3 L_{\tilde{M}}\right)\bigg]
\nonumber\\
&& + C_F T
\bigg[\frac{109}{3} - 16 L_{\tilde{M}}+12  L_{\tilde{M}}^2 +12 \frac{X_t}{\tilde{M}}\left(-1+L_{\tilde{M}}\right)
\bigg]
\nonumber\\
&& 
+ C_F C_A \bigg[-\frac{23}{72}-\frac{37}{6} L_{\tilde{M}}-\frac{1}{2}
  L_{\tilde{M}}^2 - \frac{X_t}{\tilde{M}}\left(1+3 L_{\tilde{M}}\right)
\bigg]
\bigg\}
\end{eqnarray}

\subsubsection{Scenario B}
The second scenario we consider is the so called "split-SUSY'' one, with
squarks much heavier than all the other particles:
\[
\begin{split}
&m_{\tilde{u}} = \ldots = m_{\tilde{b}} = m_{\tilde{t}} =\tilde{M}\quad \gg\quad m_{\tilde{g}}\quad \gg\quad
m_t\nonumber\\ 
& \alpha_s^{(6)} = \zeta_{s}^{\tilde{q}}\, \alpha_s^{\rm{ (SQCD)}} \,,\qquad 
m_t^{(6)} = \zeta_{m_t}^{\tilde{q}}\, m_t^{\rm{ (SQCD)}}\,.
\end{split}
\]

The result reads:
\begin{align}
\zeta_{m_t}^{\tilde{q}} =& \frac{\alpha_s}{4\pi} C_F \bigg\{ \frac{1}{2} - L_{\tilde{M}} + \frac{1}{\tilde{M}^2}(M_{\tilde{g}}^2 - 2 M_{\tilde{g}} X_t) \nonumber\\
 &+ \frac{1}{\tilde{M}^4}\bigg[ M_{\tilde{g}}^4(1 + L_{\tilde{M}}  
 - L_{\tilde{g}}) + 2 X_t M_{\tilde{g}}^3 (-1 + L_{\tilde{g}} - L_{\tilde{M}}) \bigg]\bigg\}\nonumber \displaybreak[1] \\
& + \left(\frac{\alpha_s}{4\pi}\right)^2 C_F\bigg\{-\frac{C_A}{72 \tilde{M}^4}\bigg[\tilde{M}^4\big(-481 + 432 L_{\tilde{M}} + 108 L_{\tilde{M}}^2 \nonumber \\
&  + 120 L_{\tilde{g}} - 72 L_{\tilde{g}}^2 + 576 \zeta(2)\big) - 36 M_{\tilde{g}}^3\big(M_{\tilde{g}}(38 + 4 L_{\tilde{M}} \nonumber \\
& + 13 L_{\tilde{M}}^2 + 2 L_{\tilde{g}} - 20 L_{\tilde{M}}L_{\tilde{g}} + 7L_{\tilde{g}}^2 - 20 \zeta(2)) \nonumber \\
& - 4 X_t(5 + 19 L_{\tilde{M}} + 3 L_{\tilde{M}}^2 - 16L_{\tilde{g}} - 3 L_{\tilde{M}} L_{\tilde{g}} + 2 \zeta(2))\big) \nonumber \\
& + 72 \tilde{M}^2 M_{\tilde{g}}\big(2 X_t(7 + 6L_{\tilde{M}} - 3 L_{\tilde{g}} + 2 \zeta(2))\nonumber\\
& + M_{\tilde{g}}(-15 + 3 L_{\tilde{M}} - 6 L_{\tilde{g}} + 10 \zeta(2))\big)\bigg]\nonumber \displaybreak[1]\\
& +\frac{C_F}{8\tilde{M}^4}\bigg[ -8 \tilde{M}^2 M_{\tilde{g}}\big(M_{\tilde{g}}(21 + L_{\tilde{M}} - 20\zeta(2)) \nonumber \\
& + X_t(5- 6L_{\tilde{M}} - 8\zeta(2))\big) - 2 M_{\tilde{g}}^3 \big(M_{\tilde{g}}(175 + 60 L_{\tilde{M}}^2 \nonumber \\
& + 90 L_{\tilde{g}} + 56 L_{\tilde{g}}^2 - 2L_{\tilde{M}}(43 + 58 L_{\tilde{g}}) - 104 \zeta(2)) \nonumber\\
& + 4 X_t (15 - 6 L_{\tilde{M}}^2 + 11 L_{\tilde{g}} + L_{\tilde{M}}(-17 + 6L_{\tilde{g}}) - 8 \zeta(2))\big)\nonumber \\
& + \tilde{M}^4\big(-189 - 48 L_{\tilde{M}} + 4 L_{\tilde{M}}^2 + 120 \zeta(2)\big)\bigg] \nonumber\displaybreak[1] \\
& + \frac{T}{3 \tilde{M}^4}\bigg[-36 \tilde{M}^2 M_{\tilde{g}} \big((-1 + L_{\tilde{M}})M_{\tilde{g}} + X_t - 2 L_{\tilde{M}} X_t\big) \nonumber \\
& + 3 M_{\tilde{g}}^3\big((-13 - 12 L_{\tilde{M}}^2 + 6L_{\tilde{g}} + 6L_{\tilde{M}}(-3 + 2 L_{\tilde{g}}))M_{\tilde{g}} \nonumber \\
& + 4 (7 + 6 L_{\tilde{M}}^2 + L_{\tilde{M}}(9 - 6L_{\tilde{g}}) - 3 L_{\tilde{g}})X_t\big) \nonumber \\
& + \tilde{M}^4\big(127 - 30 L_{\tilde{M}} + 36 L_{\tilde{M}}^2 - 36 \zeta(2)\big)\bigg]\bigg\}
\end{align}
We have checked the formulae above against the exact calculation both
analytically and numerically. First, we have computed the
two-loop decoupling coefficient $\zeta_{m_t}$ for scenarios A and B making use
of asymptotic expansion method (explicitly Large Mass expansion) that is
available 
in the code {\tt exp} and compared with the expansion of the exact
result. For scenario A we obtain agreement for the dominant term ({\it i.e.}
neglecting corrections proportional with the mass differences between the SUSY
particles.) For scenario B, we verified the agreement for the first three
terms of the expansion in mass ratio $M_{\tilde g}^2/\tilde{M}^2$. Also, the
direct numerical comparison of the exact and asymptotically expanded 
results    gives very good agreement. 

\subsection{\label{sec:numerics} Numerical results}
\begin{figure}[t]
  \begin{center}
      \epsfig{figure=\figurepath 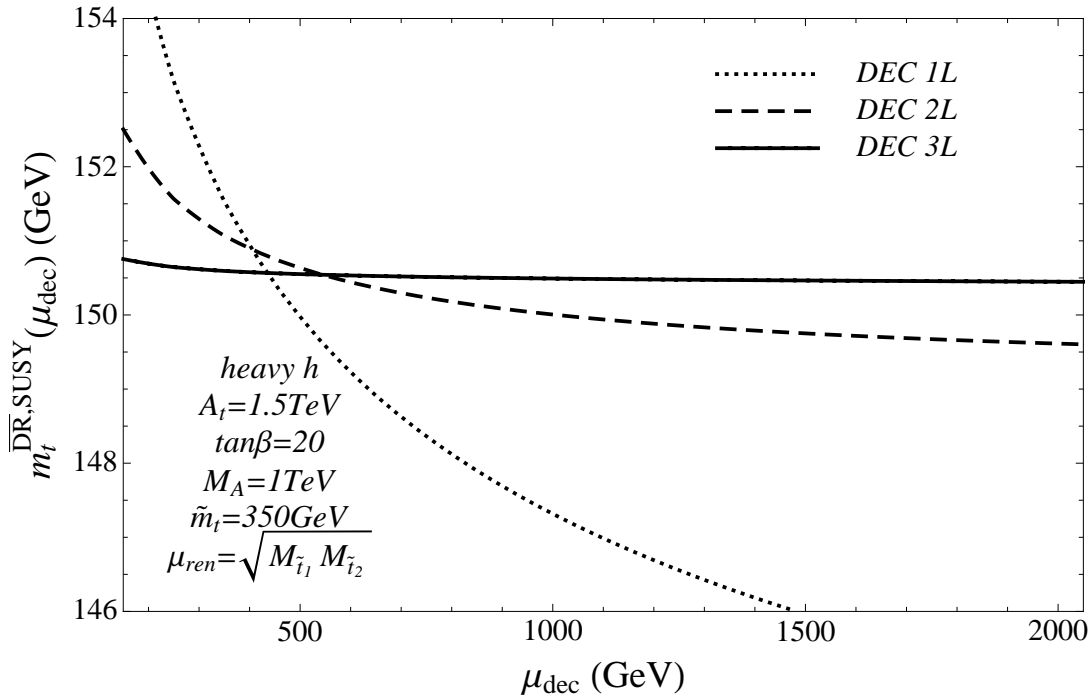,width=.9\textwidth}
      \caption[]{\label{fig:mudec} Decoupling scale dependence of
        the running top quark mass for the ``heavy Higgs'' scenario. The
         curves show the results obtained within the decoupling method at
        one (dotted line), two (dashed line) and three  (full line) loops. The
        renormalization scale was fixed to $\mu_{\rm ren}=\sqrt{M_{\tilde{t}_1}M_{\tilde{t}_2}}$.
        }
  \end{center}

\end{figure}

In this section we present the phenomenological effects of the two-loop SUSY-QCD
threshold corrections on the prediction of the running top-quark mass at the
SUSY mass scale. We also present the comparison with the direct prediction
obtained from the ratio between the running and the pole mass within SUSY-QCD as
described in the code {\tt TSIL}~\cite{Martin:2005qm}.\\
Our method can be summarised in the following sequence:
\begin{equation}
\begin{split}
M_t^{\rm OS}\,\stackrel{(i)}{\to}\,
m_t^{\msbarmath}(M_t)\,\stackrel{(ii)}{\to}\,
m_t^{\msbarmath}(\mu_{\rm dec})\,\stackrel{(iii)}{\to}\,
m_t^{\rm{ SQCD},\overline{\rm DR}}(\mu_{\rm dec})\,\stackrel{(iv)}{\to}\,
m_t^{\rm{ SQCD},\overline{\rm DR}}(\mu)\,,
\label{eq::mbrun}
\end{split}
\end{equation}
where $M_t^{\rm OS}$ denotes the top-quark pole mass and $m_t^{\msbarmath}$
and $m_t^{\rm{ SQCD},\overline{\rm DR}}$ stand for the running top-quark mass in the
SM and SUSY-QCD in the \msbar{} and \drbar{} schemes, respectively. $\mu_{\rm
  dec}$ is the  scale at which the decoupling is performed 
 and it is usually chosen  comparable with SUSY masses.
If not stated otherwise, we fix it to be the arithmetic average over the squarks and gluino on-shell masses:
\begin{equation}
 \mu_{\rm{dec}}=\frac{1}{13}\big[M_{\tilde{g}}+\sum_{\tilde{q}}M_{\tilde{q}}\big]\,.
\end{equation}
But of course, $\mu_{\rm{dec}}$ can be chosen arbitrarily and the dependence
of the running top-quark on it is a measure of the theoretical
uncertainties (for details, see Fig.~\ref{fig:mudec} and its discussion).
Also in the numerical setup, we implemented it as a free
parameter that can be varied. In the step $(i)$ the relation between the top-quark  pole
 and  running masses in the SM is required. We implemented the
 three-loop results~\cite{Chetyrkin:1999ys,Chetyrkin:1999qi,Melnikov:2000qh,Marquard:2007uj}. The
 RGEs for the SM and the MSSM  that are necessary 
 in the steps $(ii)$ and $(iv)$ are known to
 three-loop
 accuracy~\cite{Chetyrkin:2012rz,Bednyakov:2012en,Ferreira:1996ug,Harlander:2009mn}
 as well. Let us mention, that in QCD the quark anomalous dimension was recently computed even at
 the five loop order~\cite{Baikov:2014qja}. For consistency, the threshold 
 corrections evaluated in the step $(iii)$ are necessary at two loops.

For the explicit numerical evaluation we use  for the SM parameters   their
values cited in Ref.~\cite{pdg}. For the MSSM parameters we employ two
scenarios that we call "heavy Higgs'' and "heavy sfermions'', respectively. We
obtain the numerical values with the help of the spectrum generator {\tt
  SOFTSUSY}~\cite{Allanach:2001kg}. The input parameters for the spectrum
generator are as follows:

i) In the "heavy sfermions'' scenario we define all \drbar{} breaking parameters at the input scale following the Supersymmetry Les Houches Accord (SLHA)~\cite{Skands:2003cj,Allanach:2008qq} \verb|EXTPAR 0| $Q_{\rm{in}}=\tilde{m}_t$.
Where $\tilde{m}_t$ is an alias for the right handed top squark mass breaking parameter \verb|EXTPAR 46| and kept as free parameter.
Further we identify the third generation doublet mass breaking $\tilde{m}_{Q_3}$ alias \verb|EXTPAR 43| with $\tilde{m}_t$.
All other sfermion mass breaking parameters have a common value $\tilde{m}_f=\tilde{m}_t+1\rm{TeV}$.
The trilinear couplings (\verb|EXTPAR 11-13|) are given by $A_{t}=20\rm{GeV}$ and $A_{\tau}=A_b=4\rm{TeV}$.
The gaugino mass parameters (\verb|EXTPAR 1-3|) are set to $M_1=M_2=M_3=1.5 \rm{TeV}$.
For the bilinear coupling of $H_u$ and $H_d$ (\verb|EXTPAR 23|) we chose $\mu_{\rm{SUSY}}=200\rm{GeV}$.
The mass of the pseudo scalar Higgs boson (\verb|EXTPAR 23|) is set to $M_A=1\rm{TeV}$.
Finally, the ratio of the two vacuum expectation values (\verb|MINPAR 3|) is set to $\tan\beta=10$. 
The given parameters choice results in very weakly mixing top squarks, 
which are about $1\rm{TeV}$ lighter than the other sfermions and thus have largest impact in the decoupling process.
By increasing the value of $\tilde{m}_t$ one automatically pushes the squark mass spectrum to higher values.
However, in order to successfully describe the currently measured mass for the lightest Higgs boson,
one is forced to use  multi TeV range values for $\tilde{m}_t$, because of the
very weak mixing between the top squarks which have nearly equal masses.

ii) In the "heavy Higgs'' scenario we define all \drbar{} breaking parameters at the input scale $Q_{\rm{in}}=\sqrt{M_{\rm{SUSY}}^2+M_t^{\rm{OS}\,2}}$.
Here $M_{\rm{SUSY}}=1\rm{TeV}$ is a common breaking mass parameter for all sfermions except $\tilde{m}_t$ (\verb|EXTPAR 46|), which we keep as free parameter.
The remaining input parameters are given by:
\begin{align}
 A_{\tau}&=A_{b}=2469.49{\rm{GeV}}, & A_t&=1.5{\rm{TeV}}, &\mu_{\rm{SUSY}}&=200\rm{GeV},\nonumber\\
 M_1&= 5 s^2_w/(3c_w^2)M_2,& M_2 &=200\rm{GeV},& M_3 &=800\rm{GeV},\nonumber\\
 M_A&=1{\rm{TeV}}, & \tan\beta&=20\,. & &
\end{align}
Here $s_w$ and $c_w$ are the sine and cosine of the Weinberg mixing angle $\theta_W$.
In contrast to the first scenario we do have light Higgs masses for sub TeV values of $\tilde{m}_t$ due to the stop mixing.
Moreover one can have a very light stop of order $300{\rm{GeV}}$ for $\tilde{m}_t$ values having nearly same size.

Please note that for the pure SUSY-QCD analysis done in this paper, 
the effect of changing breaking parameters of particles transforming as QCD
singlets is very weak with respect to SUSY-QCD decoupling effects. We provide
the SLHA input files for the two scenarios considered here in electronic
format on the web page http://www.ttp.kit.edu/Progdata/ttp14/ttp14-025.

In the following, we denote as leading order (LO) value for the running
top-quark mass in the decoupling method
(DEC), the value obtained with one-loop RGEs and without threshold
corrections. The next-to-leading (NLO) and the next-to-next-to-leading order
(NNLO) values in the decoupling method are calculated employing two- and
three-loop RGEs and one- and two-loop threshold corrections, respectively. The
NLO top-quark mass computed in the direct method ( with the code {\tt TSIL}) takes
into account the one-loop relation between the running- and pole-quark mass,
whereas the NNLO prediction is based on the two-loop relation.

In a first step, we study the dependence of the running top-quark mass on the
unphysical parameter $\mu_{\rm{dec}}$. As this parameter is not fixed by the
theory, the dependence of the physical quantities on it provides us a measure
of the accuracy of the method  itself. We display in Fig.~\ref{fig:mudec} such
a dependence for the ``heavy Higgs'' scenario. It is particularly important
to perform the study in this scenario as one of the supersymmetric particles (
the light top-squark) has the mass at an intermediate scale of about
$300$~GeV, whereas the rest of the particles have masses around $1$~TeV. The natural question
to be addressed in this case is whether the one-step decoupling approach, where 
 all supersymmetric particles are integrated out at once, is still appropriate or a
multi-step procedure is required. In the   Fig.~\ref{fig:mudec} the dotted,
dashed and full lines depict the running top quark mass evaluated at scale
$\mu_{\rm ren}=\sqrt{M_{\tilde{t}_1}M_{\tilde{t}_2}}$ at LO, NLO and NNLO and the decoupling scale is
varied in the range from $M_{\rm top}$ to $2 M_{\rm SUSY}$. As expected from
theoretical consideration and clearly illustrated in the
plot,  a substantial improvement of the stability of the  predictions
w.r.t. the variation of $\mu_{\rm{dec}}$ is obtained when going from one-
to three-loop accuracy. While at the two-loop level a variation of the
top-quark mass of about $2$~GeV is still present, at  three loops the
variation amounts to less than $100$~MeV.  Since the expected experimental
accuracy for the top-quark mass measurement even at the future ILC does not go below
$100$~MeV, we can conclude that the method proposed here is well suited also
 for scenarios with lighter supersymmetric particles with masses around $300$~GeV.

\begin{figure}
  \begin{center}
      \epsfig{figure=\figurepath 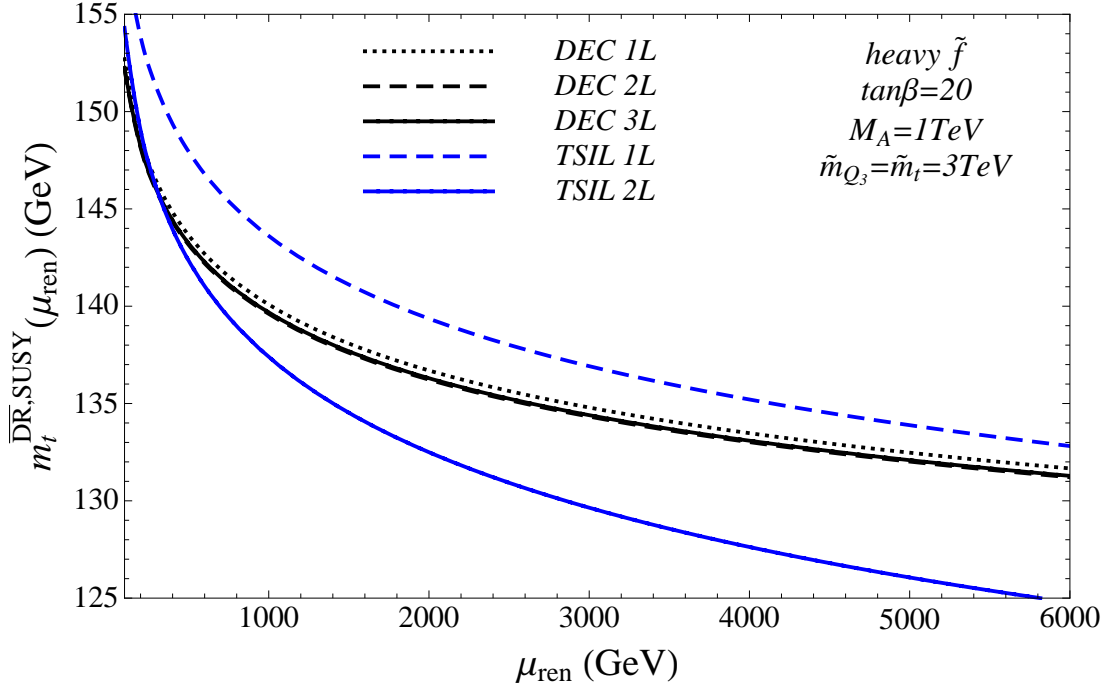,width=.91\textwidth}
      \caption[]{\label{fig:mtrenhsf} Renormalization scale dependence of
        the running top quark mass for the ``heavy sfermion'' scenario. The
        black curves show the results obtained within the decoupling method at
        one (dotted line), two (dashed line) and three (full line) loops. The
        blue lines display the results obtained with the code {\tt TSIL} at
        one (dashed line) and two (full line) loops.   
        }
  \end{center}

\end{figure}

Furthermore, in Fig.~\ref{fig:mtrenhsf} we present the running top-quark mass in the full
theory (in the ``heavy sfermion'' scenario) as a function of the scale $\mu_{\rm ren}$ at which it is evaluated. 
The black (middle)  curves  display the results obtained with the method
proposed in this paper (that we refer at as the decoupling method
and is denoted as ``DEC'' in the legend of the
plot)
 at one (dotted line) two (dashed line) and three
(full line) loops. The blue curves show the predictions obtained directly via the
ratio between the top-quark pole  and the running masses in the SUSY-QCD,
using the code {\tt TSIL}. The dashed line corresponds to the
one-loop results and the full line displays the two-loop contributions.  As
can be seen from the figure, the radiative corrections for the decoupling method
are very small (tenth of MeV between one- and two-loop order contributions and
negligible at the three loops) as
compared with the current experimental uncertainty on the 
top quark pole mass of about $1$~GeV. The perturbative series is very
well converging and the contributions from the unknown higher order
corrections are negligible for all renormalization scales. The 
 radiative corrections obtained via the direct method are much larger than
the experimental uncertainty. Even at the two-loop order, they amount to
$10$~GeV at the electro-weak scale and increase further at renormalization
scales comparable with the squark masses. In this case higher order
contributions are necessary to bring the theoretical precision at the same
level as the experimental one. One observes also that the predictions obtained
in the two methods at the two-loop order agree very well for small renormalization scales below
$400$~GeV. This can be explained by the fact that in this case the
 logarithmic contributions (of the form $\ln(M_{\rm top}/\mu)$) are small and
 the resummation gives only minor 
corrections. When the running top-quark mass is evaluated at high energy
scales the resummation of the large logarithms becomes important and the
difference between the two predictions can reach about $10$~GeV. Let us also
point out that in the domain where the resummation is expected to bring only small 
effects the differences between the two  methods decrease considerably when
going from one to two loops as is expected in perturbation theory.

\begin{figure}
  \begin{center}
      \epsfig{figure=\figurepath 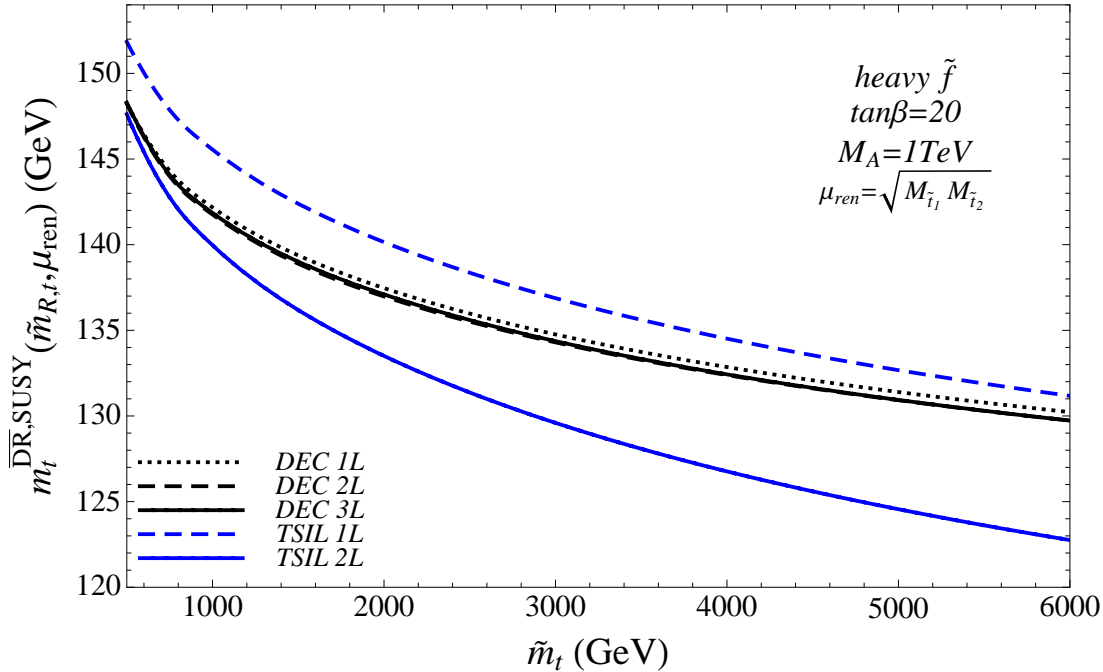,width=.91\textwidth}
      \caption[]{\label{fig:mtmrhsf} Dependence of
        the running top-quark mass on the SUSY   scale $\tilde{m}_{t}$
        for the ``heavy sfermion'' scenario.  The convention for the curves is
        the same as in Fig.~\ref{fig:mtrenhsf}.          
        }
  \end{center}

\end{figure}

In Fig.~\ref{fig:mtmrhsf} the running top-quark mass is shown as a function of
the  squark mass breaking parameter $\tilde{m}_{t}$, that can be interpreted
in the ``heavy sfermion'' scenario as a scale for the SUSY mass spectrum. In this case the renormalization scale
is chosen as geometric mean value of the top-squark masses. One notices that
the radiative corrections calculated with the decoupling approach are very small for all
SUSY scales. The direct computation deliver even at the two-loop order
radiative  corrections of about ${\cal O}(10)$~GeV. As expected, the two
methods provide results in very good agreement at the two-loop order for low
SUSY mass scales, where no large logarithmic corrections are present.

\begin{figure}
  \begin{center}
      \epsfig{figure=\figurepath 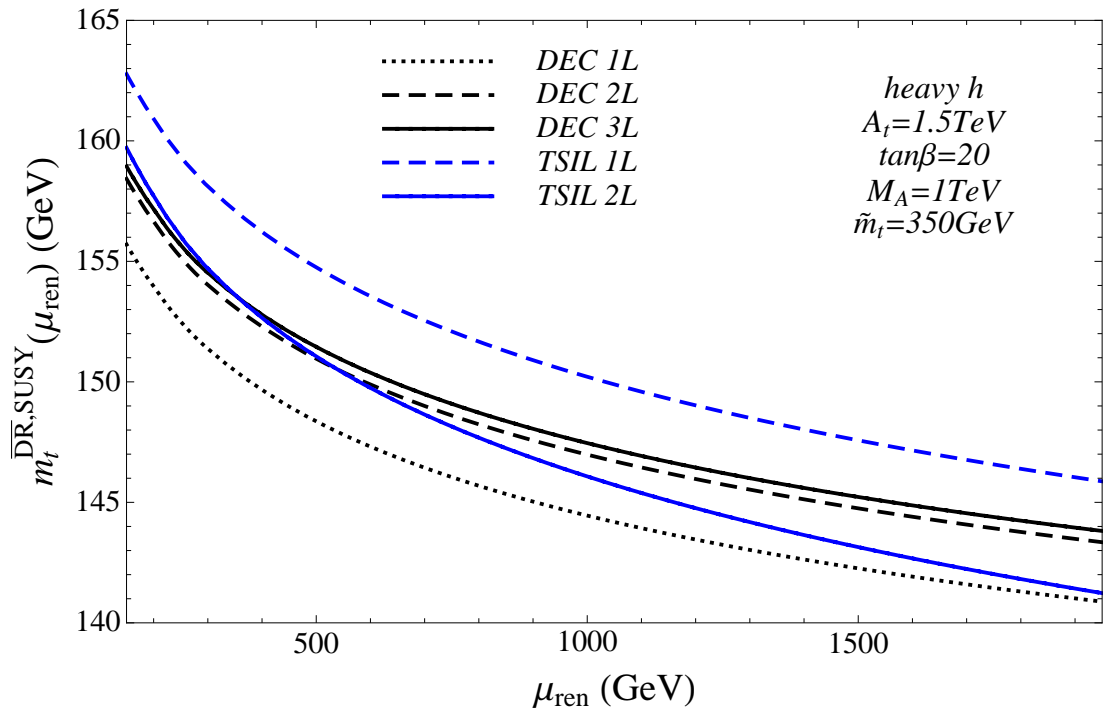,width=.91\textwidth}
      \caption[]{\label{fig:mthh}  Dependence of  the running top quark mass
      on the  renormalization scale within  the ``heavy Higgs'' scenario. The convention for the curves is
        the same as in Fig.~\ref{fig:mtrenhsf}.      
        }
  \end{center}

\end{figure}

Fig.~\ref{fig:mthh} shows the running top-quark mass as  a function of
the renormalization scale in the "heavy Higgs'' scenario. For the chosen
input parameters, the predictions of the two methods differ at the one-loop
level by more than $10$~GeV. At the two-loop level the predictions of the two
approaches agree well within the present experimental uncertainty on the
top-quark pole mass for low-energy scales, whereas for large renormalization scales the difference
amounts to  few GeV. Let us also point out that the differences between one- and
two-loop contributions within the decoupling method amounts to  about $4$~GeV,
whereas the genuine three-loop contributions are very small, below
$100$~MeV. This observation proves the good convergence of the perturbative
methods in the  decoupling approach. The direct calculation based on the code
{\tt TSIL} provides similarly
 large radiative corrections at the two-loop level. However, in order to
 reduce the theoretical uncertainty at a similar level with the experimental
 one, we  need  higher order radiative corrections  that are
 currently not available.

In summary, we conclude that the two methods provide results in good agreement
for low SUSY mass scales or renormalization scales, but they differ
significantly when the SUSY particles become heavy, in the multi TeV
range. Also, the rapid convergence of the perturbative series for the decoupling
method allows us to reduce the genuine theoretical uncertainties due to unknown
higher order corrections well below the present experimental uncertainty on the
top-quark pole mass. The discrepancies between the predictions obtained within
the two methods can have important phenomenological implications, depending on
the process and observables under consideration.

\section{\label{sect:mhphen} The mass of the lightest Higgs boson}
The  Higgs boson mass  measurement by ATLAS $125.5\pm 0.2\pm
 0.6$~GeV~\cite{ATLAS:2013mma} and 
 CMS $125.7\pm 0.3\pm 0.3 $~GeV~\cite{CMS:yva} already reached   an
 amazing precision. In the MSSM, the
  lightest Higgs boson mass is predicted. 
 Beyond the tree-level approximation, it is a function of the top-squark
 masses and mixing parameters. It  grows logarithmically with the 
top-squark masses and can be used to determine an upper bound for the
supersymmetric (SUSY) mass scale from the measured Higgs boson mass, once the mixing
parameters are fixed. This approach has received considerable attention
recently~\cite{Feng:2013tvd,Hahn:2013ria,Draper:2013oza}, partially because
the direct searches for SUSY particles 
at the LHC remained unsuccessful, indicating a  possible lower bound for their mass
scale in the  TeV range.\\
Since the dependence of the Higgs boson mass  on the SUSY  masses is
 logarithmic, high-precision measurements and theoretical predictions are
 required. The genuine theoretical
 uncertainties due to unknown higher order corrections are expected to
 grow with the SUSY mass scale. For top-squark masses in the multi TeV
 range  they were estimated to be of about
 $5-7$~GeV~\cite{Hahn:2013ria,Draper:2013oza}. 
By now, the complete one-loop~\cite{Ellis:1990nz,Brignole:1992uf} and
dominant two-loop~\cite{Heinemeyer:1998np,Degrassi:2001yf}
corrections are implemented 
in the numerical  programs FeynHiggs~\cite{Heinemeyer:1998yj} and
CPsuperH~\cite{Lee:2003nta} using on-shell 
particle masses, and in SOFTSUSY~\cite{Allanach:2001kg} and SPheno~\cite{Porod:2003um} using
running parameters. The 
dominant SUSY-QCD  three-loop corrections are taken into account in the code
{\tt H3m}~\cite{Kant:2010tf}, for which a mixed renormalization scheme was
employed. However, the three-loop contributions are known only for specific
mass hierarchies between the SUSY particles.  
 The dominant contributions to the  leading (LL) and 
next-to-leading (NLL)
logarithmic terms in the ratio between the top quark mass and 
 the typical scale of SUSY particle masses $\ln(M_{\rm
   top}/M_{\rm SUSY})$ have been obtained in
 Ref\cite{Martin:2007pg}. Very recently, the generalization of the LL and NLL
 approximation  has been derived~\cite{Hahn:2013ria} and implemented
 in the  code {\tt FeynHiggs}, up
 to the seventh loop-order. Furthermore, in
 Ref.~\cite{Draper:2013oza} the recent calculations of the 
 three-loop beta-functions for the SM  
 coupling
 and the two-loop corrections to the Higgs boson mass in 
 the SM
 have been used to derive (presumably) the dominant NNLL  corrections at the four-loop  order.

In this section we focus on the numerical effects that the  new prescription
for the determination of the running top-quark mass and top-Yukawa coupling
has on the predictions of the lightest Higgs boson mass taking into account
SUSY-QCD radiative corrections. We implemented the
 resummation of the large logarithms of the form $\ln(M_{\rm   top}/M_{\rm SUSY})$ contained in
  the running of the top-Yukawa coupling (as discussed in the previous
  section) on top of the three-loop SUSY-QCD corrections to the lightest
  Higgs  boson mass encoded in the 
 program~{\tt  H3m}. As will be shown, this type of resummation is necessary  for SUSY
  masses in the multi TeV range and enable us to  reduce  the effects
 of the  unknown higher order contributions. In the following, we evaluate the
 running top-quark mass and couplings with the highest accuracy both for the
 decoupling and the direct methods and use them further for the calculation of
 the lightest Higgs boson mass at one-, two- and three-loop accuracy. 
 
\begin{figure}
  \begin{center}
      \epsfig{figure=\figurepath 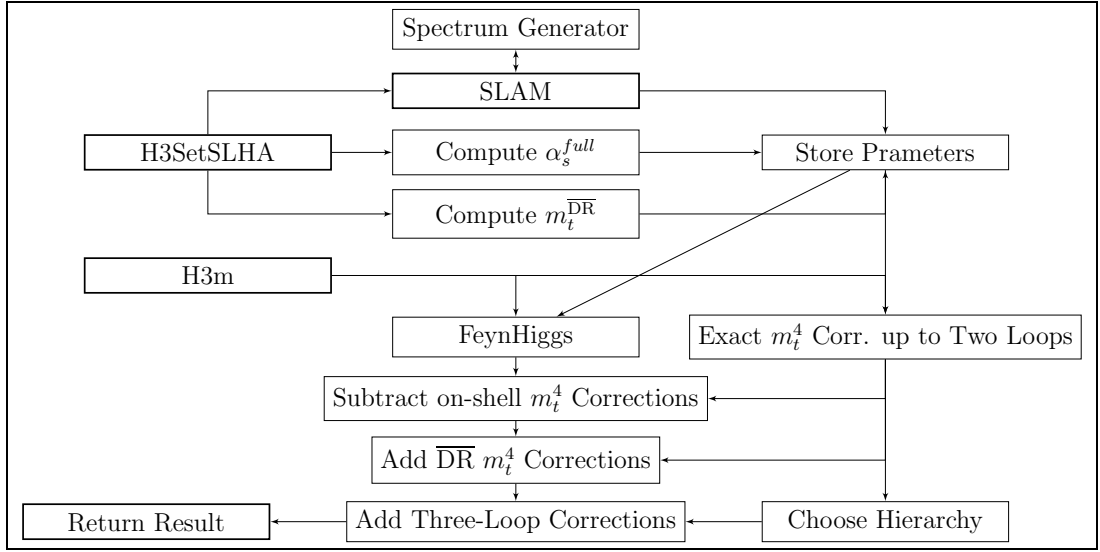,width=.91\textwidth}
      \caption[]{\label{fig:flow} Flowchart of the new version of {\tt H3m}. First, the user calls
the code {\tt SLAM} to set the MSSM  parameters. A subsequent call of {\tt H3m}
computes $M_h$.}
  \end{center}
\end{figure}

 In order to allow  a convenient evaluation of general MSSM scenarios
 including the stated low scale ones, 
 the {\tt Mathematica} package {\tt SLAM}~\cite{Marquard:2013ita} has been implemented in {\tt  H3m}.
 It provides an easy to use interface for calling and reading SUSY
 spectrum generator output full automatically using the SLHA. 
 Moreover it enables the ability to save and recall SUSY spectra in and from a data base.
 Besides reading in user provided SLHA spectrum generator output files, 
 it is now possible to enter the SLHA input file for the spectrum generators
 defining the SUSY scenario directly in {\tt Mathematica}  
 relaxing the restriction to predefined scenarios in earlier versions of {\tt
   H3m}.\\
The MSSM parameters derived in this way are stored and further used to compute the running
top-quark mass, the running top-Yukawa coupling and the strong coupling
constant in the MSSM using the decoupling method described in
section~\ref{sec:frm}. This step is realized through stand-alone routines as
is also explained in the flowchart plot in Fig.~\ref{fig:flow}. Afterwards, the
stored values for the input MSSM parameters together with  the \drbar{}
couplings just derived are delivered to the code {\tt H3m}. The computation  
 of the lightest Higgs boson mass follows then the  steps described in
 Ref.~\cite{Kant:2010tf}. The user has also the possibility to choose the way
 the running top-quark mass is computed.
    The command\\ {\tt SetOptions[H3GetSLHA, calcmt->\{``MtTSIL''\}]} allows to use
 at this stage  the code  {\tt TSIL}, whereas the decoupling method is
 implemented as default option. A
 direct comparison between the predictions obtained with the two methods will
 be presented below.\\
 The new version of the code {\tt H3m} together with few simple example
 programs are available from
{\tt  https://www.ttp.kit.edu/Progdata/ttp14/ttp14-025/}. Apart from the
implementation of the interface program {\tt SLAM} and the routine for the
computation of the running top-quark mass and the strong coupling constant 
through the decoupling method, we improved on the determination of the 
mixing angle and reduced oscillations in the Higgs mass by setting the W-boson mass
fixed to is on-shell value as given by the PDG~\cite{pdg}. 


\begin{figure}
  \begin{center}
      \epsfig{figure=\figurepath 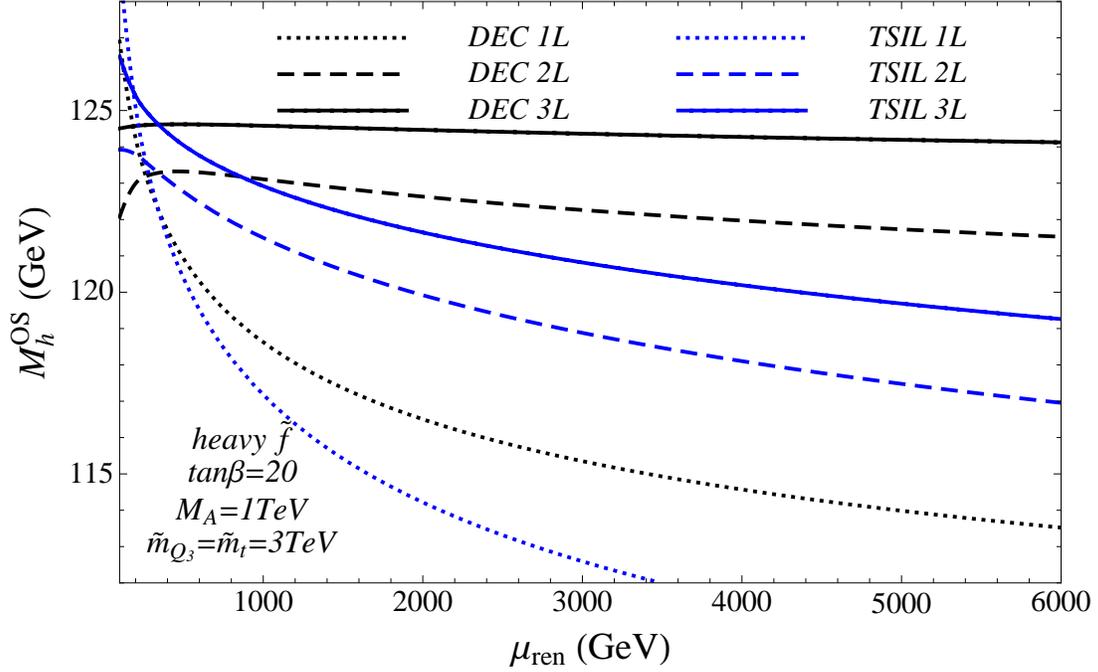,width=.9\textwidth}
      \caption[]{\label{fig:mhrenhsf} Dependence of  the Higgs boson mass
      on the  renormalization scale within  the ``heavy sfermion'' scenario.
      The black curves display the one-(dotted line), two-(dashed line) and
      three-loop (full line) 
      contributions to the lightest Higgs boson mass, when the running
      top-quark mass is determined using the decoupling method. For all
      three curves the running top-quark mass is evaluated with three-loop
      accuracy. The blue curves present the same predictions using for the
      calculation of the   running top-quark mass the code {\tt TSIL}. Beyond
      one-loop, only the SUSY-QCD contributions are taken into consideration.
        }
  \end{center}
\end{figure}

In Fig.~\ref{fig:mhrenhsf} it is shown the dependence  of the lightest Higgs
boson mass $M_h$ in the ``heavy sfermion'' scenario on the renormalization
scale, taking into account beyond 
one-loop only SUSY-QCD radiative corrections. The black curves  display the
one-(dotted line), two-(dashed line) and 
      three-loop (full line)       contributions obtained with the running
      top-quark mass in the decoupling method. The blue curves present the
      same predictions using the code {\tt TSIL} for the 
      calculation of the   running top-quark mass. The
      explicit value of the running-top quark mass  can be read from
      Fig.~\ref{fig:mtrenhsf}. It is
      known that the renormalization scale dependence of an observable gives
      an estimation for the magnitude of unknown higher order corrections.
      This enables us to use it for the determination of the theoretical uncertainty.
       As expected the renormalization scale dependence is reduced when going
       from one- to two- and to three-loop order corrections in both
       schemes. However, the direct determination of the running top-quark mass
       (blue curves) is affected by large logarithmic corrections that in turn
       induces large radiative corrections to the Higgs boson mass. Even at
       the three-loop order, the scale variation of $M_h$
       amounts to about $5$~GeV, more than an order of magnitude larger than
       the current
       experimental accuracy on $M_h$ and few times bigger than the parametric
       uncertainties. In contrast, the resummation of the   logarithmic
       corrections to the running top-Yukawa coupling through the decoupling
       method renders the scale dependence of $M_h$ at three-loop order very
       mild about tens of MeV. One can also observe, that  low values for the 
       renormalization scale around the top-quark pole mass are not well
       suited for the present scenario, especially when the three-loop order
       contributions are not taken into account. In this case, radiative
       corrections even beyond the three-loop order are required in order to
       cope with the experimental precision. The difference between the
       predictions for $M_h$ obtained with the two methods is sizeable and can
       amount to few GeV for large values of the renormalization
       scale.

\begin{figure}
  \begin{center}
      \epsfig{figure=\figurepath 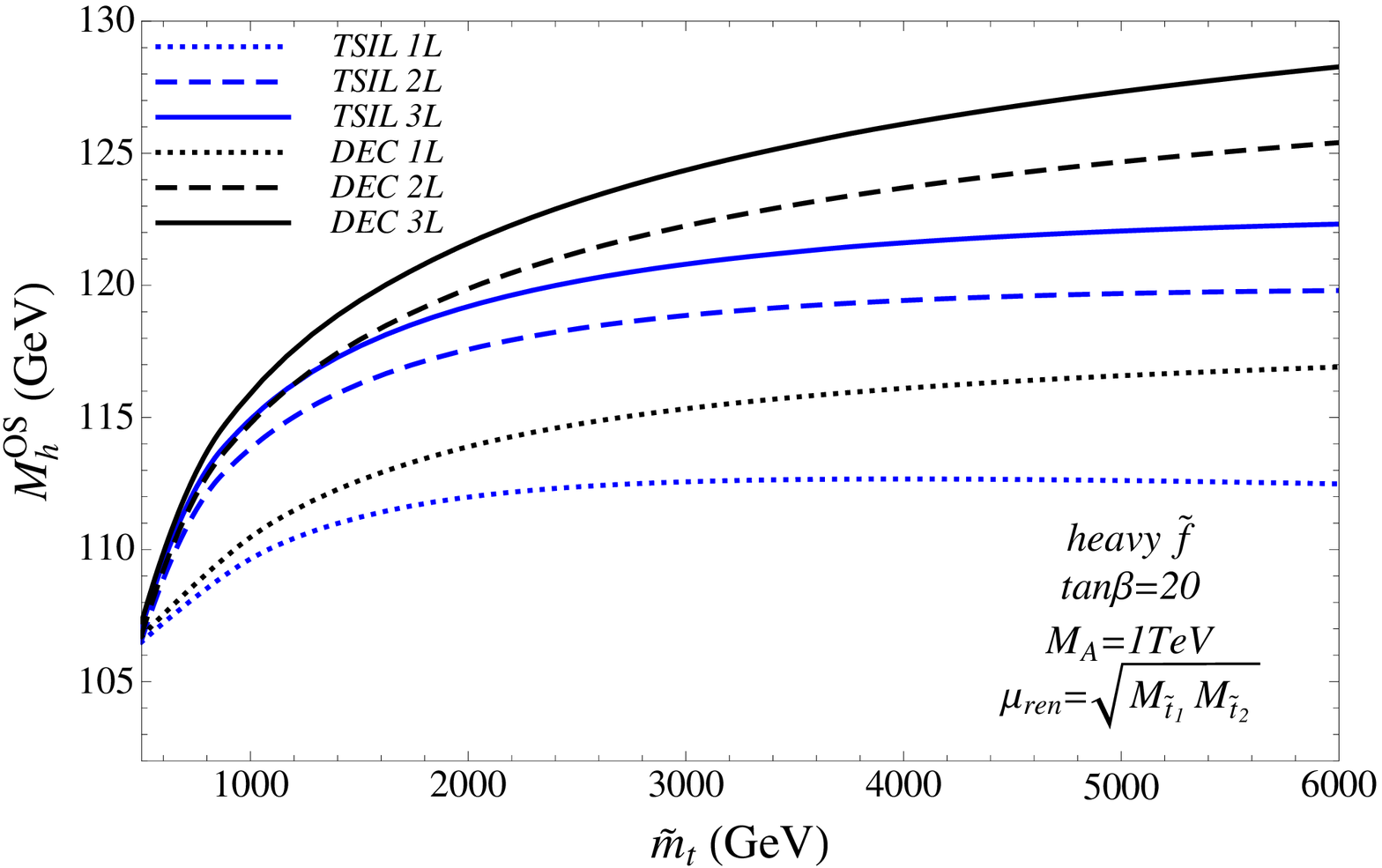,width=.91\textwidth}
      \caption[]{\label{fig:mhmrhsf} Dependence of
        the lightest Higgs boson mass on the SUSY  scale $\tilde{m}_{t}$
        for the ``heavy sfermion'' scenario.  The same convention for the line
        style is used as for the Fig.~\ref{fig:mhrenhsf}.        
        }
  \end{center}
\end{figure}

The dependence of $M_h$ on the SUSY breaking parameter $\tilde{m}_{t}$ is
shown in Fig.~\ref{fig:mhmrhsf}. As described in section~\ref{sec:numerics},
$\tilde{m}_{t}$ can be interpreted as an estimation of the SUSY mass
parameter. The convention for the line style is the same as in the previous
figure.  The renormalization scale is fixed as the geometric mean value of the top-squark
masses, thus in the TeV range. The radiative
corrections to the Higgs boson mass increase with the SUSY mass scale as
expected. The predictions obtained using the two
methods for the derivation of the top-Yukawa coupling are in good agreement
for low SUSY scales of about $500$~GeV, but they differ significantly 
for heavy SUSY spectrum in the multi TeV range. The Higgs boson mass predicted
through the decoupling method is always  heavier  and has a steeper dependence
on the SUSY spectrum as compared to the one obtained through
the direct method. This difference can be explained by the effects of
resumming large logarithms within the first approach. Since we use the RGEs at
the three-loop order the next-to-next-to-leading-logarithms  are resummed. A
similar behaviour of predictions 
for $M_h$ based on resummation was observed in the previous
works~\cite{Hahn:2013ria,Draper:2013oza}. Let us point out the big impact of
the resummation procedure for constraining the SUSY parameter space. While the
prediction of $M_h$ through the decoupling method allows SUSY mass scales of
about $4$~TeV, the present scenario is already excluded when the direct method
is employed. This observation explains also the need for very precise theoretical
predictions for the Higgs boson mass.

\begin{figure}
  \begin{center}
      \epsfig{figure=\figurepath 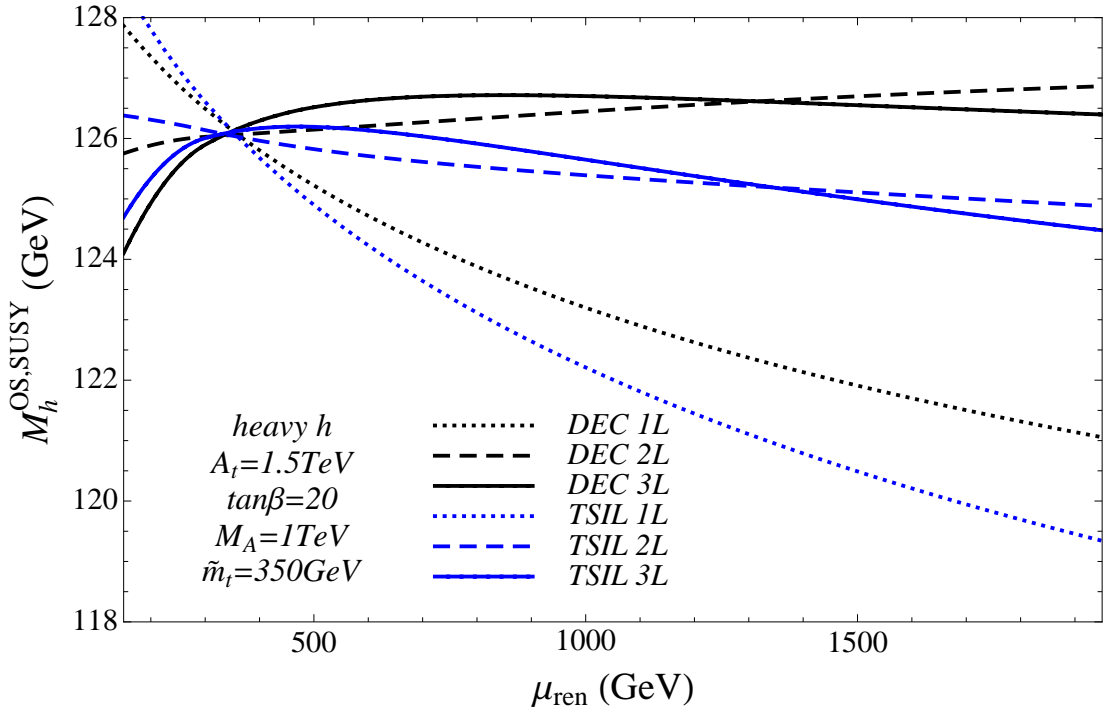,width=.91\textwidth}
      \caption[]{\label{fig:mhrenhh}  Dependence of  the lightest Higgs boson mass
      on the  renormalization scale within  the ``heavy Higgs'' scenario.
      The same convention for the   line style is used as in Fig.~\ref{fig:mhrenhsf}.  
        }
  \end{center}
\end{figure}

In Fig.~\ref{fig:mhrenhh} the dependence of $M_h$ on the renormalization scale
is shown for the ``heavy Higgs'' scenario. As can be read from  the figure, 
the scale dependence is  reduced when higher order radiative corrections are
taken into account. However, even at the three-loop order the variation of
$M_h$ with the renormalization scale can amount to few GeV for both
methods. It is also important to notice that low values of the 
renormalization scale are characterized by large radiative contributions. A
better alternative is the choice of the renormalization scales  above $500$~GeV. In the
decoupling method this choice will reduce  the scale variation to about
$200$~MeV. As for the previous scenario, we observe a milder scale variation
when the decoupling method is employed, that can be  associated with smaller
higher order corrections.  The difference between the predictions obtained in
the two frameworks amounts to few GeV for renormalization scales in the TeV
range. The two methods provide same values for $M_h$ for renormalization scale
of about $400$~MeV, for which also the predictions for the running top-quark mass coincide.

\section{\label{sec:concl} Conclusions}
In this paper, we consider the calculation of ${\cal
  O}(\alpha_s^2)$ radiative corrections to the running top-quark mass and
Yukawa coupling with the MSSM. Our method is based on the ``running and
decoupling'' technique, that has the advantage to resumm the large logarithms
by the use of RGEs. Our numerical analysis performed in
section~\ref{sec:numerics} showed that the method is very stable upon higher
order radiative corrections. The remaining theoretical uncertainty  is estimated
by half of the magnitude of genuine three-loop order contributions and amounts
to about $100$~MeV. The method proposed here provides results for the running
top-quark mass in good agreement with the 
predictions of the code {\tt TSIL} for light SUSY spectra and for low-energy
scales at which the top-quark mass is computed. For heavy SUSY particles and
for high renormalization scales the differences between the two methods can
easily reach few GeV. However, the relation between the running and the on-shell
top-quark mass in the MSSM, on which the code {\tt TSIL} rely, is affected by
large radiative corrections in the range of GeV. One also observes that the
two methods agree better when 
going from LO to NLO and NNLO.\\
The exact analytical results of our method are available in electronic
format. In addition,  we provide in the paper the results for two specific
mass hierarchies.

The second part of the paper presents  the
 effects that the new determination of the running top-Yukawa
coupling has on the prediction of the lightest Higgs boson mass in the SUSY-QCD. 
We observe a much milder dependence on the renormalization scale  of the Higgs
boson mass predicted with three-loop accuracy. The renormalization scale
dependence  is usually interpreted as a measure for the missing  higher 
order corrections. Thus, this improvement is  very welcomed given the present
difficulty to achieve  radiative corrections to the  Higgs boson mass beyond
the two-loop level. We also notice that the predictions obtained through the
decoupling method are  higher than the one derived in the direct
method. The difference can amount to several GeV for SUSY masses in the TeV
range. This behaviour can be explained by the effect of resumming the large
logarithms through the use of RGEs in the determination of the running
top-Yukawa coupling.\\
Furthermore, we implemented the  decoupling method described above together
with  the code {\tt SLAM}, that provides an 
interface  to  spectrum generators,  in the existing code {\tt H3m}.  In this
way, the code {\tt H3m} computes the 
three-loop SUSY-QCD corrections to the Higgs boson mass, taking into account
the resummation of the large logarithms of the form $\ln(M_{\rm   top}/M_{\rm SUSY})$.  

Finally, we want to stress that for the final prediction of the lightest Higgs
boson mass within the MSSM, in addition to the SUSY-QCD corrections discussed
in this paper also higher order corrections induced by the top- and
bottom-Yukawa couplings have to be considered. A similar analysis for the 
Yukawa sector is not yet feasible because the two-loop contributions to the
decoupling coefficient of the Yukawa couplings induced by mixed QCD and top- or
bottom-Yukawa corrections are not known in the literature. This analysis is
beyond the scope of the present paper and we postpone it to a future project. 
However, it is important to emphasize that the contributions that we do not
consider here (including 
two-loop top- and bottom-Yukawa  and electro-weak corrections) can
range from $0.5$~GeV to $2.5$~GeV without resummation effects, depending
on the parameters of the two 
scenarios analysed. These numbers have to be
compared with the sum of the one- and  two- and three-loop SUSY-QCD
contributions  that can reach from $20$~GeV up to $35$~GeV when no
resummation effects are included. Thus, 
the effects of  resummation discussed above that account only for the
SUSY-QCD sector, 
although not complete, provide an essential  contribution to the prediction of
the lightest Higgs boson mass for supersymmetric models with heavy particles.


\bigskip
\noindent
{\large\bf Acknowledgments}\\ 
We are grateful to Matthias Steinhauser for continuous support and for
enlightening conversations. We thank Philipp Kant for providing
us assistance with the code {\tt H3m} and for many useful discussions. 
This work was supported by the DFG through SFB/TR~9 ``Computational Particle
Physics''.



\end{document}